\documentclass[preprint,final,12pt]{elsarticle}

\usepackage{hyperref}
\usepackage[utf8]{inputenc} 
\usepackage{graphicx}
\usepackage{subfigure}
\usepackage{amsmath}
\usepackage{adjustbox}						
\usepackage{color}
\usepackage[normalem]{ulem}

\usepackage{lipsum}

\newcommand\blfootnote[1]{%
  \begingroup
  \renewcommand\thefootnote{}\footnote{#1}%
  \addtocounter{footnote}{-1}%
  \endgroup
}

\journal{Expert Systems with Applications}

\biboptions{authoryear}

\begin{document}

\begin{frontmatter}

\title{End-to-End Environmental Sound Classification using a 1D Convolutional Neural Network}

\author{Sajjad Abdoli\fnref{myfootnote}}\ead{sajjad.abdoli.1@ens.etsmtl.ca}
\author{Patrick Cardinal}\ead{patrick.cardinal@etsmtl.ca}
\author{Alessandro Lameiras Koerich}\ead{alessandro.koerich@etsmtl.ca}

\address{Department of Software and IT Engineering, \'{E}cole de Technologie Sup\'{e}rieure, Universit\'{e} du Qu\'{e}bec, H3C 1K, Montreal, QC, Canada}
\fntext[myfootnote]{Corresponding author}

\begin{abstract}
In this paper, we present an end-to-end approach for environmental sound classification based on a 1D Convolution Neural Network (CNN) that learns a representation directly from the audio signal. Several convolutional layers are used to capture the signal's fine time structure and learn diverse filters that are relevant to the classification task. The proposed approach can deal with audio signals of any length as it splits the signal into overlapped frames using a sliding window. Different architectures considering several input sizes are evaluated, including the initialization of the first convolutional layer with a Gammatone filterbank that models the human auditory filter response in the cochlea. The performance of the proposed end-to-end approach in classifying environmental sounds was assessed on the UrbanSound8k dataset and the experimental results have shown that it achieves 89\% of mean accuracy. Therefore, the propose approach outperforms most of the state-of-the-art approaches that use handcrafted features or 2D representations as input. Furthermore, the proposed approach has a small number of parameters compared to other architectures found in the literature, which reduces the amount of data required for training.\blfootnote{ Abbreviations: AI: Air conditioner; BC: Between Class; CA: Car horn; CH: Children playing; CNN: Convolution Neural Network; DO: Dog bark; DR: Drilling; EN: Engine; GU: Gun shot; JA: Jackhammer; SI: Siren; SKM: Spherical K-Means; ST: Street music; SVMs: Support Vector Machines.}
\end{abstract}

\end{frontmatter}

\section{Introduction}
\label{sec:intro}
In the last years, Convolutional Neural Networks (CNNs) have had significant impact on several audio and music processing tasks such as automatic music tagging \citep{dieleman2014end}, large-scale video clip classification based on audio information \citep{hershey2017cnn}, music genre classification \citep{Costa2017}, speaker identification \citep{Ravanelli2018}, environmental sound classification \citep{piczak2015environmental,2017deepsalamon,pons2018randomly,simonyan2014very, tokozume2017learning} , among others. Environmental sound classification is an interesting problem \citep{sigtia2016automaticreview,stowell2015detectionreview} which has different applications ranging from crime detection \citep{radhakrishnan2005audiocrime} to environmental context aware processing \citep{chu2009environmental}. Moreover, with the increasing interest in smart cities, IOT devices embedding automatic audio classification can be very useful for urban acoustic monitoring  \citep{mydlarz2017urbanmonitoring} like intelligent audio-based surveillance system in public transportation \citep{laffitte2019assessing}.

Like typical automatic classification systems, most of the approaches for environmental sound classification rely on handcrafted features or learn representations from mid-level representations such as spectro-temporal features \citep{ludena2016acoustic,Costa2012}. Spectral representations have been used as features in several approaches based on matrix factorization \citep{mesaros2015soundmatrixf,benetos2016detectionmatrixf,bisot2016acousticmatrixf,salamon2015unsupervised,geiger2015gaborGMM}.
\citet{mesaros2015soundmatrixf} presented an approach for overlapping sound event detection based on learning non-negative dictionaries through joint use of spectrum and class activity annotation. \citet{benetos2016detectionmatrixf} presented an approach for overlapping acoustic event detection based on probabilistic latent component analysis where each exemplar in a sound event dictionary consists of a succession of spectral templates. \citet{bisot2016acousticmatrixf} learn features from time-frequency images in an unsupervised manner. The images are decomposed using matrix factorization methods to build a dictionary and the projection coefficients are used as features for classification. \citet{salamon2015unsupervised} proposed a dictionary learning method based on the Spherical K-Means (SKM) algorithm which used log-Mel spectrograms as input. \citet{geiger2015gaborGMM} used Gabor filterbank features and Gaussian Mixture Models for event detection. \citet{mulimani2019segmentation} used a singular value decomposition method for extracting acoustic event specific features from spectrogram. Theses features are used as inputs to a Support Vector Machine (SVM) classifier. Recently, \citet{XIE2019} proposed a method for aggregation of acoustic and visual features for acoustic scene classification. Several acoustic features like spectral centroid, spectral entropy as well as several visual features like local binary pattern, histogram of gradients are proposed. A suitable feature selection algorithm like principle component analysis is also used. The selected feature set is used as input to an SVM classifier. 

Recent works explore CNN-based approaches given the significant improvements over hand-crafted feature-based methods \citep{piczak2015environmental,2017deepsalamon,pons2018randomly,simonyan2014very, tokozume2017learning}. However, most of these approaches first convert the audio signal into a 2D representation (spectrogram) and use 2D CNN architectures that were originally designed for object recognition such as AlexNet and VGG \citep{simonyan2014very}. One of the main advantages of using 2D representations is that spectrograms can summarize high dimensional waveforms into a compact representation. Furthermore, 1D representations are noisier compared to 2D representations \citep{stowell2014}. Piczak \citep{piczak2015environmental} presented a CNN with two layers followed by three dense layers. The network operates on two input channels: log-Mel spectra and their deltas. However, one of the challenges in using 2D CNNs for environmental sound classification is that the modelling capacity of such networks depends on the availability of a large amount of training data to learn kernel parameters without over-fitting. The scarcity of labeled data of environmental sounds is also a problem. \citet{2017deepsalamon} presented a method based on a 2D CNN with five layers (SB-CNN) where new training samples are generated using data augmentation methods such as time stretching, pitch shifting, dynamic range compression or adding background noise \citep{mcfee2015MUDA}. The 2D CNN was trained on the augmented dataset and evaluated on the original samples. They reported the classification accuracy of 79\% on a dataset of environmental sounds \citep{Salamon:2014:DTU:2647868.2655045}. \citet{pons2018randomly} used randomly weighted 2D CNNs (non-trained) for extracting features from audio spectrograms and raw audio samples for sound classification. Several experiments have been conducted to find the best architectures for this method. In the case of environmental sound classification, the best results have been obtained by using a VGG 2D CNN \citep{simonyan2014very} as a feature extractor and SVMs as classifiers. They reported mean accuracy of 7\% for this problem. \citet{tokozume2017learning} proposed a new method called Between-Class (BC) learning for training neural networks. The network, for which the input is a mixture of two audio samples, is trained to predict the mixing ratio of the samples. According to their experiments, the BC learning has shown performance improvement for various architectures used for sound identification tasks. They also proposed an end-to-end 1D CNN (EnvNet-v2) that performs well on various environmental sound datasets when trained with the BC learning approach, compared to conventional learning techniques. The best error rate of 8.6\% is reported on ESC-10 dataset \citep{piczak2015esc}.   

1D CNNs that learn acoustic models directly from audio waveforms are becoming a popular method in audio processing due to the ability of these networks to take advantage of the signal's fine time structure \citep{hoshen2015speech}. \citet{Kim2018} proposed a 1D CNN architecture for music auto-tagging inspired by the building blocks of Resnets \citep{resnet2016} and SENets \citep{SENets}. \citet{Zhu2016} proposed an end-to-end learning approach for speech recognition based on multiscale convolutions that learns the representation directly from audio waveforms. Three 1D convolutional layers with different kernel sizes are used for feature extraction and the features are concatenated by a pooling layer for ensuring a consistent sampling frequency for the rest of the network. They reported 23.28\% of word error rate on a dataset drawn from a collection of sources including read, conversational, accented, and noisy speech. Ravanelli and Bengio \citep{Ravanelli2018} proposed the SincNet, an end-to-end approach for speaker identification and verification. The first layer of such a model is based on parametric sinc functions, which are band-pass filters. Only low and high cutoff frequencies of the filters are learned from data. This model learns meaningful filters for the first layer and decreases the number of parameters of the model. This model achieves a sentence error rate of 0.85\% on TIMIT dataset \citep{garofolo1993darpa}. \citet{Zeghidour2018} also proposed an end-to-end 1D CNN architecture for speech recognition by learning a filter bank which is considered as a replacement of Mel-filterbanks. \citet{hoshen2015speech} proposed an end-to-end multichannel 1D CNN for speech recognition. They also found that the timing difference between channels is an indicator of the location of the input in space. They reported 27.1\% of single channel word error rate on a large vocabulary voice search dataset. \citet{sainath2015learning} used a similar architecture for speech recognition. They showed that features learned directly from the audio waveform match the performance of log-Mel filterbank energies. \citet{Dai2017} proposed several very deep convolutional models for environmental sound classification that achieved 72\% of accuracy on UrbanSound8k dataset. The proposed models consist of batch normalization, residual learning, and down-sampling in the initial layers of the CNN.

In this paper, we propose an end-to-end 1D CNN for environmental sound classification that learns the representation directly from the audio waveforms instead of from 2D representations \citep{piczak2015environmental,2017deepsalamon,salamon2015unsupervised}. The proposed end-to-end approach provides a compact architecture that reduces the computation cost and the amount of data required for training. With the aim of extracting relevant information directly from audio waveforms, several convolutional layers are used to learn low-level and high-level representations. The highest level of representation is then used for classifying the input signal by means of three fully connected layers. Experimental results on UrbanSound8k dataset, which contains 8,732 environmental sounds from 10 classes, have shown that the proposed approach outperforms other approaches based on 2D representations such as spectrograms \citep{piczak2015environmental,2017deepsalamon,pons2018randomly,salamon2015unsupervised} by between 11.24\% (SB-CNN) and 27.14\% (VGG) in terms of mean accuracy. Furthermore, the proposed approach does not require data augmentation or any signal pre-processing for extracting features.

Our contribution in this paper is twofold. We present an end-to-end 1D CNN initialized with Gammatone filterbanks that has few parameters and which does not require a large amount of data for training compared to dense 2D CNNs which have millions of trainable parameters. Besides, it achieves state-of-the-art performance. Secondly, the proposed approach can handle audio signals of any length by using a sliding window of appropriate width that breaks up the audio signal into short frames of dimension compatible with the input layer of the end-to-end 1D CNN. 

This paper is organized as follows. Section \ref{sec:arch} presents the ideas behind the proposed end-to-end 1D CNN architecture and the proposed approach to deal with variable audio lengths. We also present the variations in the architecture that may arise from different input dimensions as well the process of aggregating the predictions on audio frames. Section \ref{sec:res} presents the benchmarking dataset, the experimental protocol, the procedure used to fine-tune the proposed 1D CNN to the data, the evaluation of different input sizes, the enhancements in the proposed 1D CNN to improve its performance and an analysis of the frequency response of the filters learned at the different convolutional layers. In Section \ref{sec:dis}, we compare the performance of the proposed approach with the state-of-the-art in environmental sound classification and we analyze the magnitude responses of the filters learned at the first convolutional layer to gain some insight on the behaviour of the proposed 1D CNN. Finally, the conclusions and perspectives of future work are presented in the last section. 

\section{Proposed End-to-End Architecture}
\label{sec:arch}
The aim of the proposed end-to-end architecture is to handle audio signals of variable lengths, learning directly from the audio signal, a discriminative representation that achieves a good classification performance on different environmental sounds. 

\subsection{Variable Audio Length}
\label{sub:audiolengths}
One of the challenges of using 1D CNNs in audio processing is that the length of the input sample must be fixed but the sound captured from the environment may have various duration. Therefore, it is necessary to adapt a CNN to be used with audio signals of different lengths. Moreover, a CNN must be used for continuous prediction of input audio signals of environmental sounds.

\begin{figure*}[htpb!]
  \centering
  \includegraphics[width=0.97\textwidth]{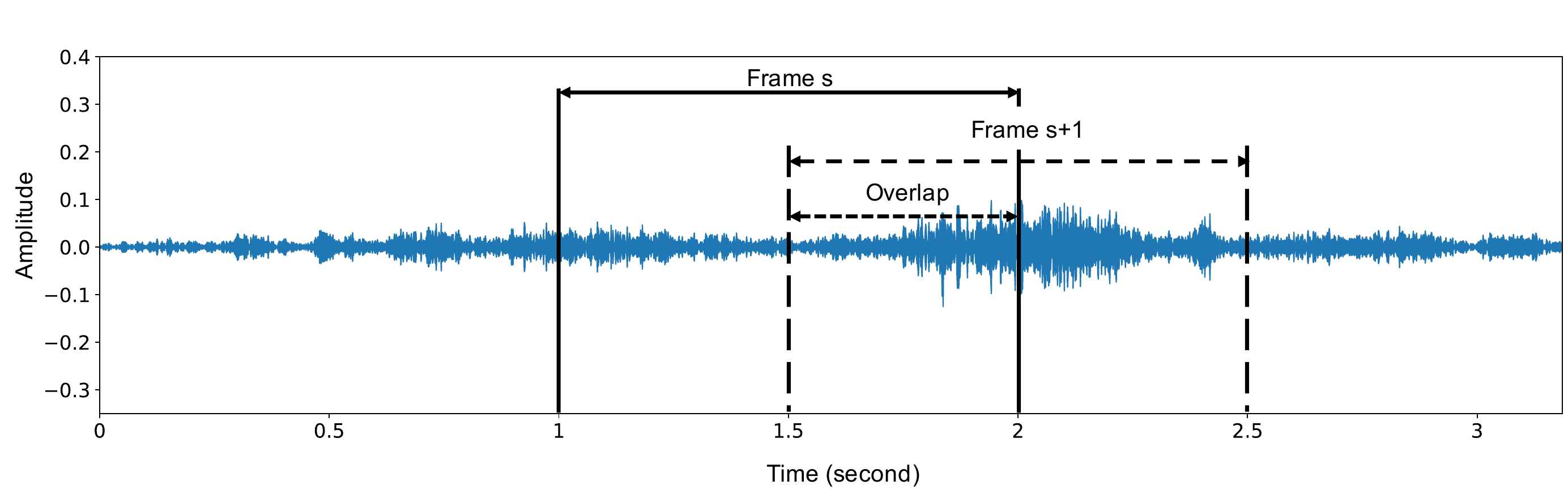}
  \caption{Framing the input audio signal into several frames ($s, s+1$) with appropriate overlapping percentage ($50\%)$.}
  \label{fig:framing}
\end{figure*}

One way to circumvent this constraint imposed by the CNN input layer is to split the audio signal into several frames of fixed length using a sliding window of appropriate width. Therefore, in our approach we use a window of variable width to conditionate the audio signal to the input layer of the proposed 1D CNN. The window width depends mainly on the signal sampling rate. Furthermore, successive audio frames may also have a certain percentage of overlapping, which aim is to maximize the use of information. This naturally increases the number of samples as some parts of the audio signal are reused and that can be viewed as some sort of data augmentation. The process of framing the audio signal into appropriate frames is illustrated in Figure \ref{fig:framing}.

Moreover, the sampling rate of the audio signals has a direct impact on the dimensionality of the input sample and eventually on the computational cost of model. For environmental sounds, a sampling rate of 16 kHz may be considered a good trade-off between the quality of the input sample and the computational cost of the model. 

\subsection{1D CNN Topology}
\label{sub:Topology}
A 1D CNN is analogous to a regular neural network but but it has generally raw data as input instead of handcrafted features. Such an input data is processed through several trainable convolutional layers for learning an appropriate representation of the input. According to the \textit{"local connectivity"} theorem, the neurons in a layer are connected only to a small region of the previous layer. This small region of connectivity is called a receptive field. The input to out 1D CNN is an array representing the audio waveform, which is denoted as \(X\). The network is designed to learn a set of parameters \( \Theta \) to map the input to the prediction \(T\) according to a hierarchical feature extraction given by Equation \ref{eq:1}:

\begin{equation}
T=F(X\mid\Theta)=f_{L}(... f_{2}(f_{1}(X\mid \Theta_{1})\mid \Theta_{2})\mid \Theta_{L})
\label{eq:1}
\end{equation}

\noindent where \(L\) is the number of hidden layers in the network. For the convolutional layers, the operation of the $l$-th layer can be expressed as:

\begin{equation}
T_{l}=f_{l}(X_{l}\mid\Theta_{l})=h(W\otimes X_{l}+b),\quad \Theta_{l}=[W,b]
\end{equation}

\noindent where \(\otimes\) denotes the convolution operation, \(X_{l}\) is a two-dimensional input matrix of \(N\) feature maps, \(W\) is a set of \(N\) one dimensional kernels (receptive field) used for extracting a new set of features from the input array, \(b\) is the bias vector, and \(h(\cdot)\) is the activation function. The shapes of \(X_{l}\), \(W\) and \(T_{l}\) are \((N,d)\), \((N,m)\) and \((N,d-m+1)\), respectively. Several pooling layers are also applied between the convolutional layers for increasing the area covered by the next receptive fields. The output of the final convolutional layer is then flattened and used as input of several stacked fully connected layers, which can be described as:

\begin{equation}
T_{l}=f_{l}(X_{l}\mid\Theta_{l})=h(W X_{l}+b),\quad \Theta_{l}=[W,b]
\end{equation}

In the case of multiclass classification, the number of neurons of the output layer is the number of classes. Using softmax as the activation function for the output layer, each output neuron indicates the membership degree of the input samples for each class. During the training process, the parameters of the network are adjusted according to the back-propagated classification error and the parameters of the network are optimized to minimize an appropriate loss function \citep{Goodfellow-et-al-2016}.

\begin{figure*}[htpb!]
  \centering
  \includegraphics[width=0.97\textwidth]{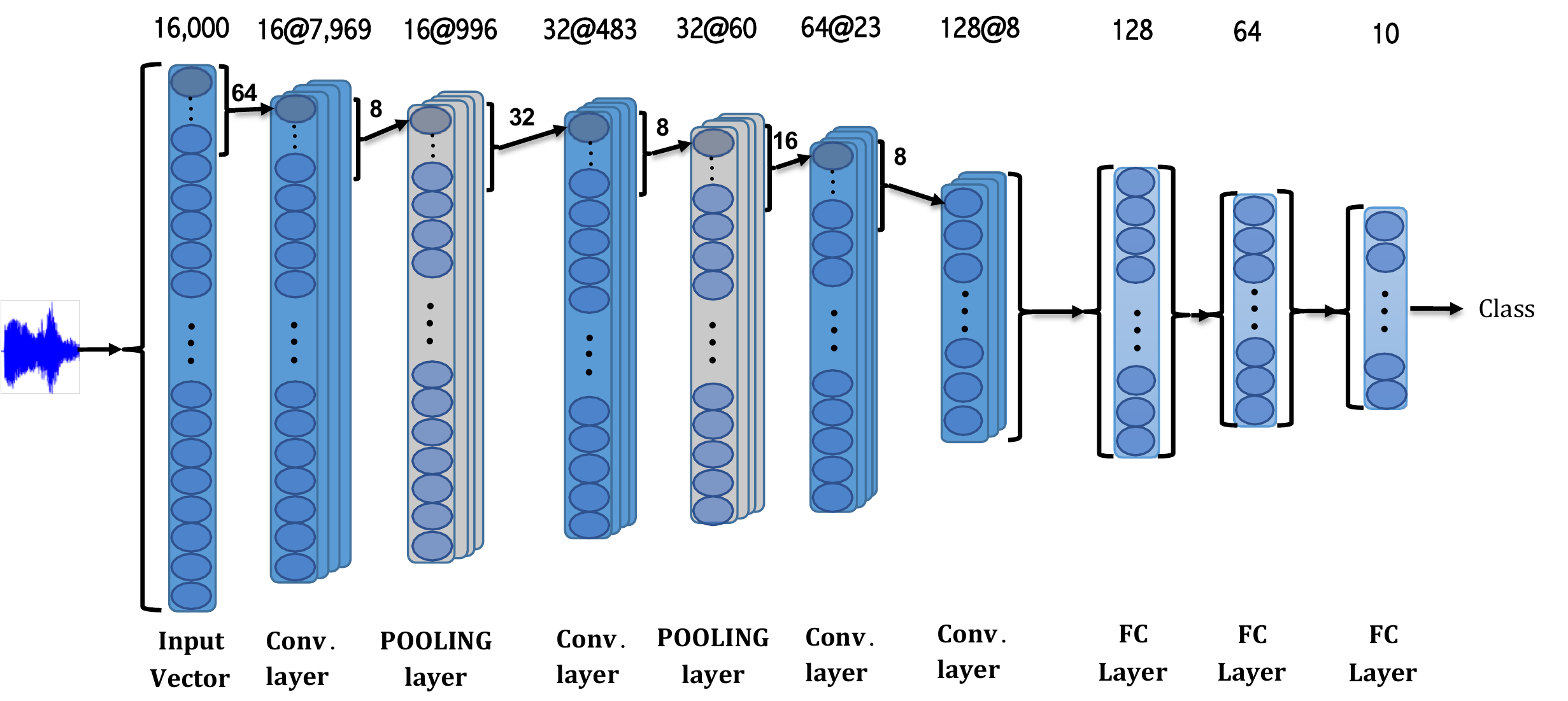}
  \caption{The architecture of the proposed end-to-end 1D CNN for environmental sound classification. The dimension, number of filters and filter size are given for the input size of 16,000. For other input sizes, the values are presented in Table \ref{tab:arch}.}
  \label{fig:architecture}
\end{figure*}

The proposed topology aims a compact 1D CNN architecture with a reduced number of parameters. The number of parameters of a CNN is directly related to the computational effort to train such a network as well as to the need of a large amount of data for training. Therefore, the proposed architecture shown in Figure \ref{fig:architecture} is made of four convolutional layers, possibly interlaced with max pooling layers, followed by two fully connected layers and an output layer. The baseline model shown in Figure \ref{fig:architecture} has as input an array of 16,000 dimensions, which represents 1-second of audio sampled at 16 kHz. However, this is not a constraint since we can adapt the model for different audio lengths and sampling rates in two ways: (i) change the model architecture to adapt it to the characteristics of the audio inputs; (ii) padding or segmenting the audio piece to adapt it to the input dimensions of the network.

Several other configurations can also be derived from subtle modifications of the base model (shown in Figure \ref{fig:architecture}) to adapt it to shorter or longer audio inputs, as shown in Table \ref{tab:arch}. This implies modifying the number of convolutional layers as well as the number and the dimension of filters and the stride. However, for long contiguous audio recordings, instead of increasing the input dimension of the network, which also implies increasing the number of parameters, and consequently its complexity, it is preferable to split the audio waveform into shorter frames by changing the window width as explained in Section \ref{sub:audiolengths}. In this way, we keep the network compact and it can process audio waveforms of any length. In spite of that, in Section \ref{sec:diffleng} we evaluate different audio lengths as input, keeping a fixed sampling rate of 16 kHz.

\begin{table*}[htpb!]
\centering
\footnotesize
\caption{The configuration of the convolutional layers (CL) and pooling layers (PL) for the end-to-end CNN considering different input sizes (audio lengths).} %
\begin{tabular}{llllllllll}
\hline\hline 
Input Size              &             & \multicolumn{8}{c}{Layer}                            \\ \cline{3-10}
                        &             & CL1    & PL1   & CL2   & PL2 & CL3 & CL4 & CL5 & PL3 \\ \hline
{50,999} & Dim         & 25,468 & 3,183 & 1,576 & 197 & 91  & 42  & 20  & 5   \\
                        & \# Filters  & 16     & 16    & 32    & 32  & 64  & 128 & 256 & 256 \\
                        & Filter Size & 64     & 8     & 32    & 8   & 16  & 8   & 4   & 4   \\
                        & Stride      & 2      & 8     & 2     & 8   & 2   & 2   & 2   & 4   \\ \hline
{32,000} & Dim         & 15,969 & 1,996 & 983   & 122 & 54  & 24  & 11  & 2   \\
                        & \# Filters  & 16     & 16    & 32    & 32  & 64  & 128 & 256 & 256 \\
                        & Filter Size & 64     & 8     & 32    & 8   & 16  & 8   & 4   & 4   \\
                        & Stride      & 2      & 8     & 2     & 8   & 2   & 2   & 2   & 4   \\ \hline
{16,000} & Dim         & 7,969  & 996   & 483   & 60  & 23  & 8   & NA & NA \\
                        & \# Filters  & 16     & 16    & 32    & 32  & 64  & 128 & NA & NA \\
                        & Filter Size & 64     & 8     & 32    & 8   & 16  & 8   & NA & NA \\
                        & Stride      & 2      & 8     & 2     & 8   & 2   & 2   & NA & NA \\ \hline
{16,000G} & Dim         & 15,489  & 19,36   & 953   & 119  & 52  & 23   & NA & NA \\
                        & \# Filters  & 64     & 64    & 32    & 32  & 64  & 128 & NA & NA \\
                        & Filter Size & 512     & 8     & 32    & 8   & 16  & 8   & NA & NA \\
                        & Stride      & 1      & 8     & 2     & 8   & 2   & 2   & NA & NA \\ \hline
{8,000}  & Dim         & 3,969  & 496   & 233   & 29  & 7   & NA & NA & NA \\
                        & \# Filters  & 16     & 16    & 32    & 32  & 64  & NA & NA & NA \\
                        & Filter Size & 64     & 8     & 32    & 8   & 16  & NA & NA & NA \\
                        & Stride      & 2      & 8     & 2     & 8   & 2   & NA & NA & NA \\ \hline
{1,600}  & Dim         & 785    & 392   & 189   & 94  & 44  & NA & NA & NA \\
                        & \# Filters  & 16     & 16    & 32    & 32  & 64  & NA & NA & NA \\
                        & Filter Size & 32     & 2     & 16    & 2   & 8   & NA & NA & NA \\
                        & Stride      & 2      & 2     & 2     & 2   & 2   & NA & NA & NA \\ \hline
\multicolumn{10}{l}{\scriptsize{NA: Not Applicable.
G: With Gammatone filterbank for the first layer of CNN.}}
\end{tabular}
\label{tab:arch} 
\end{table*}

The proposed 1D CNN has large receptive fields in the first convolutional layers since it is assumed that the first layers should have a more global view of the audio signal. Moreover, the environmental sound signal is non-stationary \textit{i.e.} the frequency or spectral contents of the signal changes with respect to time. Therefore, shorter filters do not provide a general view on the spectral contents of the signal. The output of the last pooling layer for all feature maps is flattened and used as input to a fully connected layer. In order to reduce the over-fitting, batch normalization is applied after the activation function of each convolution layer \citep{ioffe2015batch}. The last fully connected layer has ten neurons. Mean squared logarithmic error, defined in Equation \ref{eq:loss} is used as loss function ($\mathcal{L}$):

\begin{equation}
\mathcal{L}=\frac{1}{N}\sum_i^Nlog(\frac{p_{i}+1}{a_{i}+1})^{2}
\label{eq:loss}
\end{equation}

\noindent where \(p_{i}\), \(a_{i}\) and \(N\) are the predicted class, the actual class, and the number of samples respectively.

For all input sizes shown in Table \ref{tab:arch}, after the last pooling layer, there are two fully connected layers with 128 and 64 neurons respectively on which a drop-out is applied with a probability of 0.25 for both layers \citep{srivastava2014dropout}. The ReLU activation function $(h(x)=max(x,0))$ is used for all layers, except for the output layer where a softmax activation function is used. Since the amount of data for training is limited, it is not feasible to use deeper architectures without significant over-fitting. By the use of the architecture shown in Figure \ref{fig:architecture}, it is possible to omit a signal processing module because the network is powerful enough to extract relevant low-level and high-level information from the audio waveform.

The convolutional layers of the proposed architecture are inspired in \citet{aytar2016soundnet} who proposed a CNN architecture (SoundNet) for learning sound representations from unlabeled videos. The SoundNet \citep{aytar2016soundnet} learns multimodal features from audio and video using two concurrent CNNs which are further used with a SVM classifier. On the other hand, the proposed 1D CNN architecture learns the representation directly from the audio waveform, and it uses such a learned representation as input to a fully connected neural network for classification.

\subsection{Gammatone Filterbanks}
\label{sub:gamma}
Another interesting characteristic of such a 1D CNN is that its first layer can be initialized as a Gammatone filter bank. A Gammatone filter is a linear filter described by an impulse response of a gamma distribution and a sinusoidal tone. This initialization can be viewed as a trade-off between handcrafted features and representation learning. In this configuration, the kernels of the first layer are initialized by 64 band-pass Gammatone filters with central frequency ranging from 100 Hz to 8 kHz. Such a filterbank decomposes the input signal into 64 frequency bands.

\begin{figure*}[htpb!]
  \centering
  \includegraphics[width=0.97\textwidth]{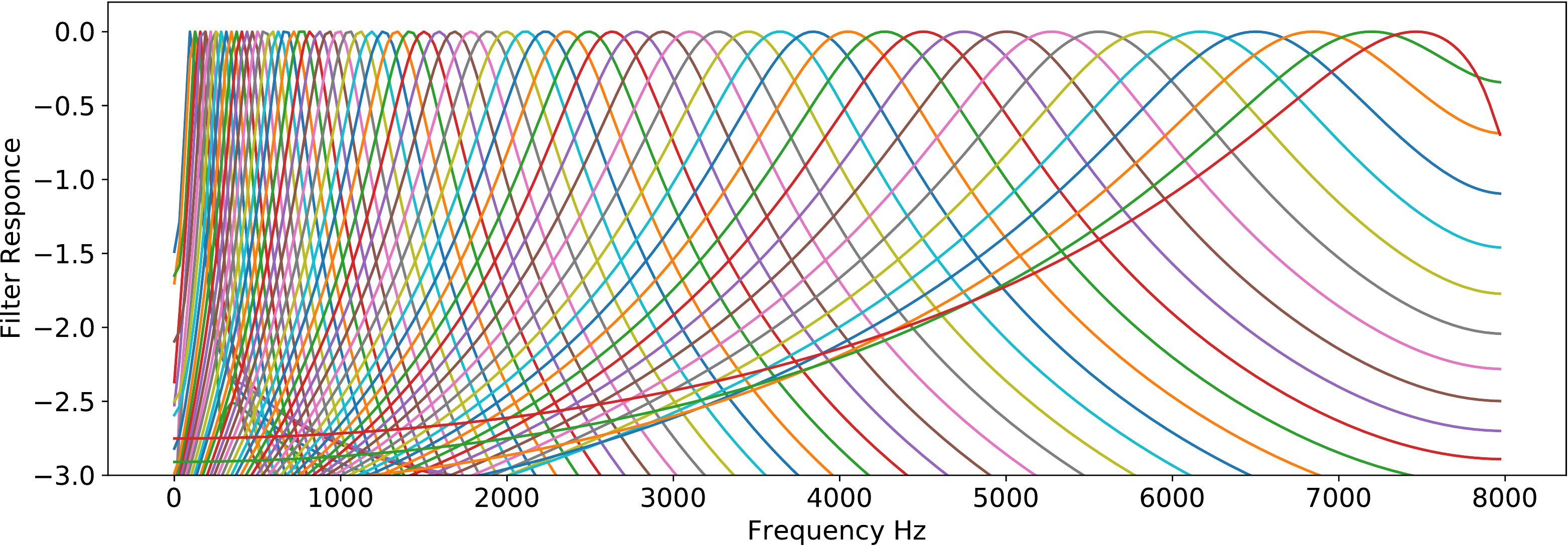}
  \caption{Frequency response of 64 filters of Gammatone filterbank.}
  \label{fig:filterbank_gamm_res}
\end{figure*}

Gammatone filters have been used in models of the human auditory system and are physiologically motivated to simulate the structure of peripheral auditory processing stage. For this reason, Gammatone filters have also been used to initialize the first layer of 1D CNNs for automatic speech recognition \citep{hoshen2015speech,Zeghidour2018,sainath2015learning}. Figure \ref{fig:filterbank_gamm_res} illustrates the frequency response of the Gammatone filterbank, generated by the Gammatone-like spectrograms toolbox developed by Ellis \citep{ellistoolbox:2009}.

\subsection{Aggregation of Audio Frames}
\label{sub:agg}
\begin{figure*}[htpb!]
  \centering
  \includegraphics[width=0.97\textwidth]{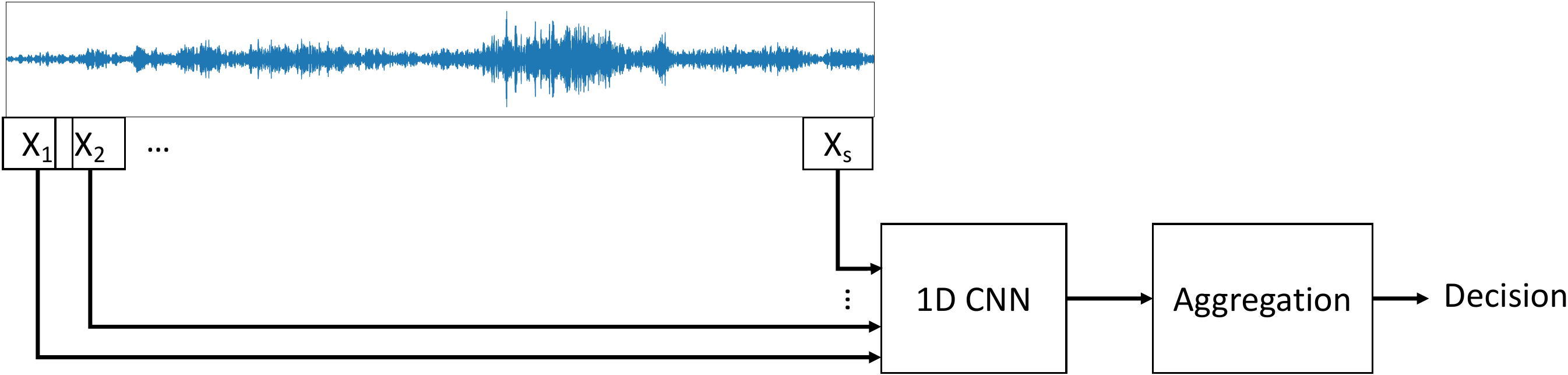}
  \caption{Aggregation of the predictions on the audio frames.}
  \label{fig:agg}
\end{figure*}

In the case where the input audio waveform $X$ is split into $S$ frames denoted as $X_1, X_2, \dots, X_S$, during the classification we need to aggregate the CNN predictions to come up to a decision on $X$, as illustrated in Figure \ref{fig:agg}. For such an aim, different fusion rules can be used to reach a final decision, such as the majority vote or the \textit{sum} rule, which are denoted in Equations \ref{eq:majvote} and \ref{eq:sumrule} respectively.

\begin{equation}
    y_i =\sum_{j=1}^S o_{ji}
    \label{eq:majvote}
\end{equation}

\noindent where $o$ is the CNN prediction for the $j=1,\dots,S$ segment of the audio waveform $X$ and $i=1,\dots,K$ is the predicted class. $S$ is the number of frames and $K$ is the number of classes.

\begin{equation}
    y_i = \frac{1}{S}\sum_{j=1}^S o_{ji}
    \label{eq:sumrule}
\end{equation}

When there are $K$ classes, we generate $K$ values and them for an audio input, we choose the class with the maximum $y_i$ value: 

\begin{equation}
    \text{ Choose \enspace}  C_i  \enspace \text{ if \enspace } y_i = \max_{k=1}^{K} y_k 
    \label{eq:decision}
\end{equation}

\section{Experimental Results}
\label{sec:res}
The proposed end-to-end 1D CNN for environmental sound classification was evaluated on the UrbanSound8k dataset \citep{Salamon:2014:DTU:2647868.2655045}. This dataset consists of 8,732 audio clips summing up to 7.3 hours of audio recordings. The maximum duration of audio clips is four seconds. The classes and the number of samples in each class are: "Air conditioner (AI): 1000", "Car horn (CA): 429", "Children playing (CH): 1000", "Dog bark (DO): 1000", "Drilling (DR): 1000", "Engine (EN) idling: 1000", "Gun shot (GU): 374", "Jackhammer (JA): 1000", "Siren (SI): 929", "Street music (ST): 1000". The original audio clips are recorded at different sample rates. For the experiments presented in this paper, they have been downsampled to 16 kHz in order to unify the shape of the input signal for the 1D CNN.  

\subsection{Fine-Tuning the 1D CNN Architecture}
The number of convolutional layers plays a key role in detecting high-level concepts. The number of convolutional layers for the base model shown in Figure \ref{fig:architecture} was determined in an exploratory experiment using the audio files of the UrbanSound8k dataset. The audio files were segmented into 16,000 samples and successive frames have 50\% of overlapping. Ten percent of the dataset was used as validation set and 10\% percent of the dataset was also used as test set. Each network was trained with 80\% of the dataset up to 100 epochs with batch sizes of 100 samples. The accuracy achieved by the 1D CNN with one to four convolutional layers on test set was 69\%, 75\%, 79\% and 80\%, respectively. Four convolutional layers is the upper limit since the minimal dimension of the feature map has been reached at this layer. The same procedure was also adopted to find the best number of convolution layers as well as their parameters for the other configurations derived from the base model which are described in Table \ref{tab:arch}.

\subsection{Evaluation on Different Audio Lengths}
\label{sec:diffleng}
All experiments reported in this subsection used a 10-fold cross-validation procedure to produce a fair comparison with the results reported by \citet{Salamon:2014:DTU:2647868.2655045}. One of the nine training folds is used as validation set for optimizing the parameters of the network to achieve the best accuracy. A batch size of 100 samples was used for training the CNNs and they were trained up to 100 epochs with early stopping. The Adadelta \citep{zeiler2012adadelta} optimizer with the default learning rate of 1.0 was used. Adadelta has been chosen because this method dynamically adapts the learning rate during the optimization process.

First, the proposed end-to-end 1D CNN is evaluated on different audio lengths to assess the impact of the input length on the classification performance. Next, the full audio recordings of UrbanSound8k dataset, which have 59,999 frames ($\approx$ three seconds), were also segmented into shorter frames using a sliding window and considering different overlapping percentages (0\%, 25\%, 50\%, and 75\%). The architecture shown in Figure \ref{fig:architecture} was adapted according to the parameters described in Table \ref{tab:arch}, leading to audio frames of 1,600 ($\approx$ 100 msec), 8,000 ($\approx$ 500 msec), 16,000 ($\approx$ 1 second) and 32,000 samples ($\approx$ 2 seconds).

The process of segmenting the audio signal into frames and aggregating the predictions of the classifier for all frames, resembles the process of aggregating the prediction of ensemble of classifiers. In this process, the most important parts of the audio signal contribute more to the final decision while the noisy or outlier frames have their importance averaged during the aggregation process. Table \ref{table:framing_acc} shows the best results achieved with different frame sizes, window overlapping and combination rules on the UrbanSound8k dataset in terms of mean accuracy. For the classification of each test sample of the original dataset, the predictions for each audio segment are combined using either the majority voting or the sum rule \citep{kittler1998combining}. Table \ref{table:framing_acc} shows that the 16,000-input architecture achieved the highest accuracy which is the same accuracy achieved by 1D CNN with 59,999 inputs, even if it has almost twice less parameters than that network. Furthermore, the 8,000-input architecture achieved a mean accuracy close to that, even if it has almost three times less parameters. On the other hand, for the 1,600-input architecture, the mean accuracy is about 7\% lower than the best architectures. This is an indication that short audio frames do not contain enough information to train properly the 1D CNN. However, this behaviour may be particular for the UrbanSound8k dataset and it cannot be generalized to other audio classification tasks or datasets.

\begin{table*}[htpb!]
\caption{Mean accuracy and standard deviation on the UrbanSound8k dataset over the 10 folds for the different architectures having as input the full audio (59,999) or segmented audio with different window widths and 50\% overlapping.}
\centering
\footnotesize
\begin{tabular}{l l c l}
\hline\hline 
\textbf{Input} & \textbf{Combination} & \textbf{Mean$\pm$SD} & \textbf{\# of } \\
\textbf{Dimension} &   \textbf{Rule} & \textbf{Accuracy} & \textbf{Parameters} \\ \hline
\multicolumn{1}{r}{59,999}                                    & NA                                    & \textbf{83\%$\pm1.3\%$} & 421,146          \\
\multicolumn{1}{r}{32,000}                                    & Maj Voting                         & 82\%$\pm0.9\%$ &  322,842         \\
\multicolumn{1}{r}{16,000}                                    & Sum Rule                                & \textbf{83\%$\pm1.3\%$} & 256,538          \\
\multicolumn{1}{r}{8,000}                                     & Sum Rule                                & 80\%$\pm1.9\%$ &  116,890        \\
\multicolumn{1}{r}{1,600}                                     & Sum Rule                                & 77\%$\pm3.0\%$ &  394,906        \\ \hline
\multicolumn{4}{l}{\scriptsize{NA: Not applicable }}
\end{tabular}
\label{table:framing_acc}
\end{table*}

\begin{figure}[htpb!]
  \centering
  \centerline{\includegraphics[width=9.5cm, height=9.5cm]{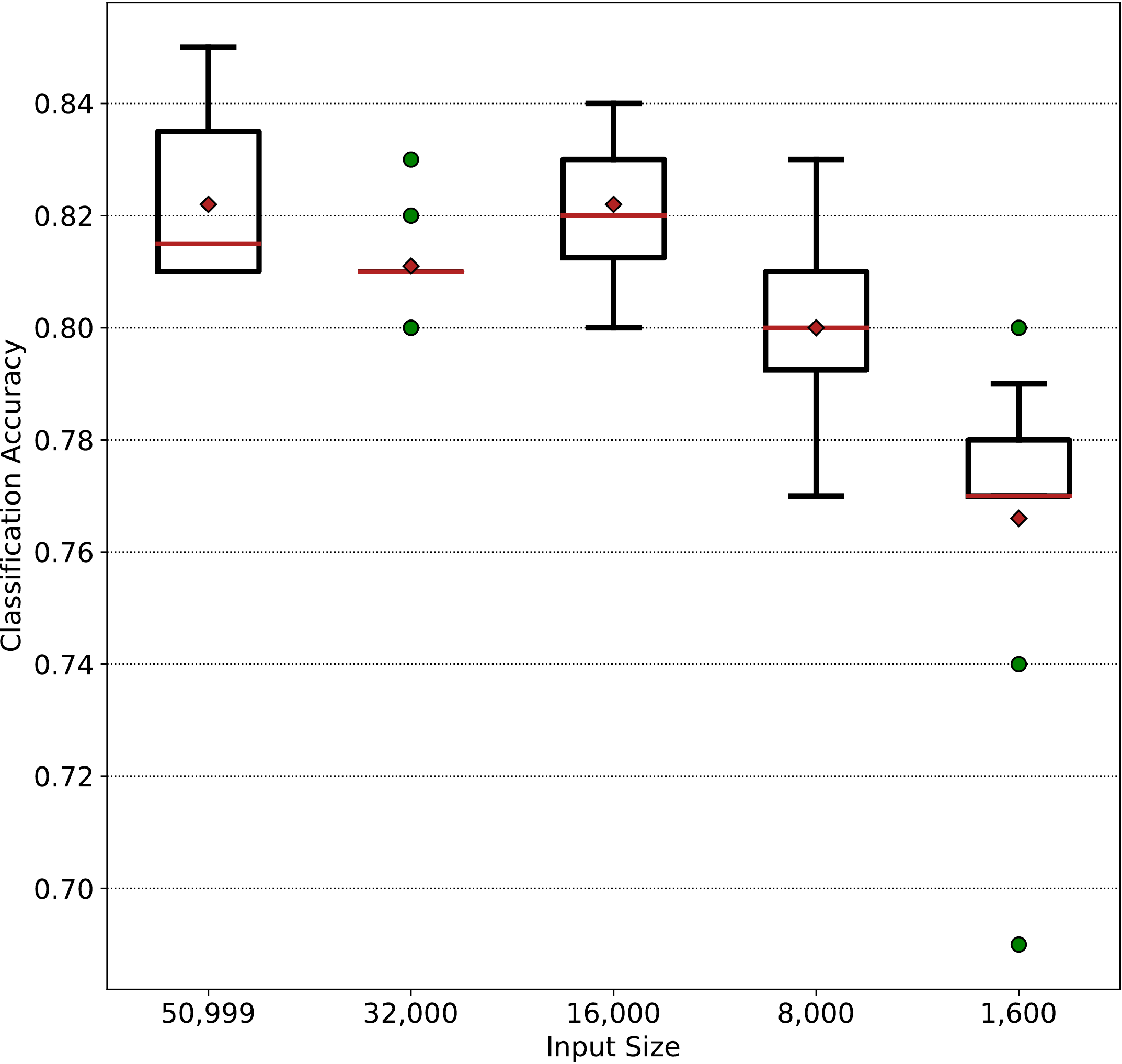}}
\caption{The box plot for the five different input sizes on UrbanSound8k dataset.}
\label{fig:boxframing}
\end{figure}

The box-plot of Figure \ref{fig:boxframing} also shows that the 16,000-input 1D CNN is the best choice since it provides the highest median; the interquartile range is the smallest one; and there is no outlier. Furthermore, such an architecture has the same mean accuracy, but almost half of the number of parameters than the second-best choice, the 50,999-input 1D CNN. Therefore, the 16,000-input 1D CNN is preferable over other architectures, as it presents the best trade-off between the number of parameters and accuracy. 

\begin{figure*}[htpb!]
  \centering
  \includegraphics[width=1\textwidth]{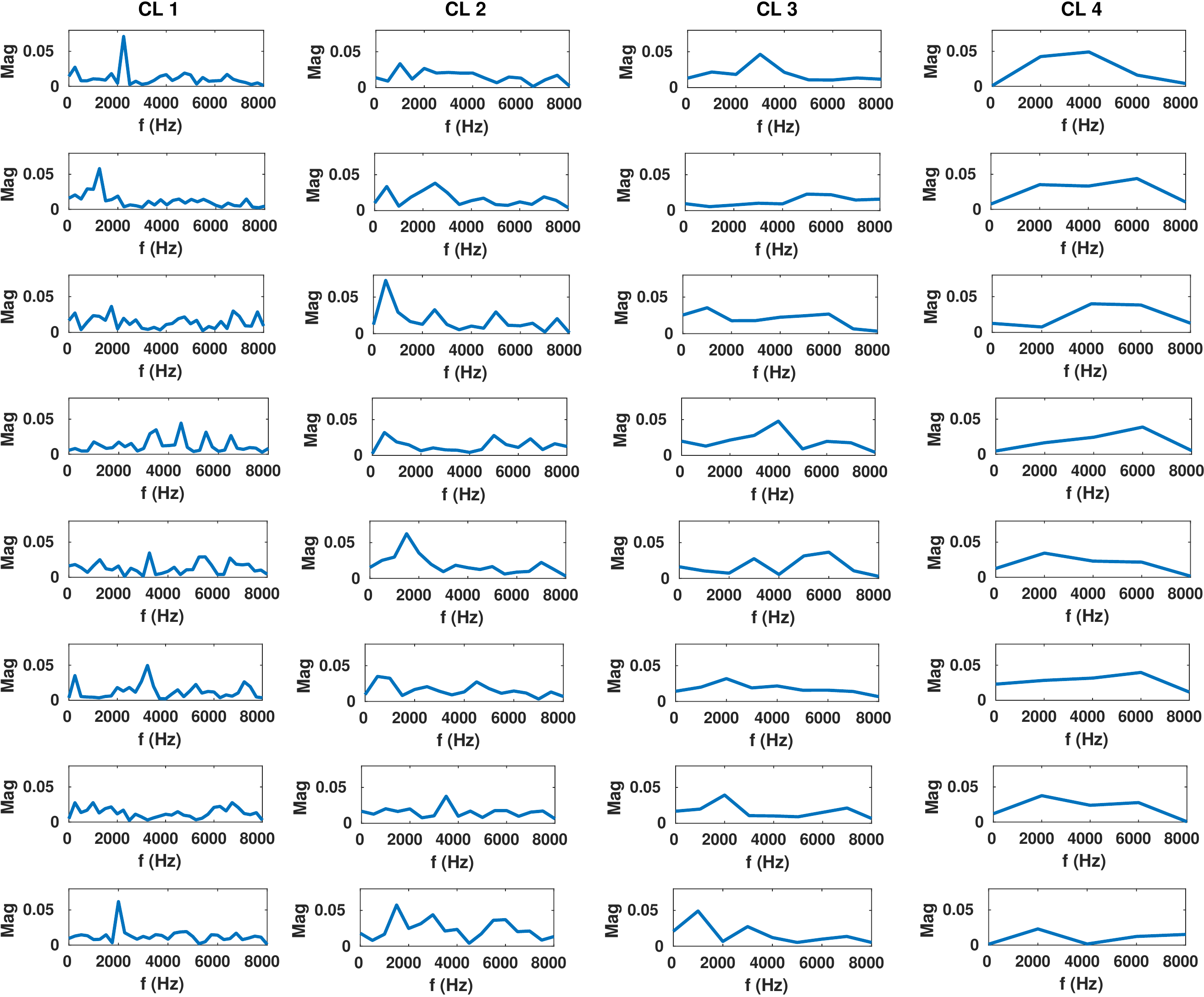}
  \caption{Fourier transform of randomly selected filters from the four convolutional layers (CLs) of the proposed 16,000-input 1D CNN shown in Figure  \ref{fig:architecture}.}
  \label{fig:kernels}
\end{figure*}

In order to have a better insight about the behavior of the convolutional filters learned by the proposed 1D CNN, the Fourier transform of some filters was computed and their frequency responses are shown in Figure \ref{fig:kernels}. These filters were randomly initialized and trained for the specific task and all of their parameters, such as central frequency, bandwidth, gain/attenuation, were learned directly from the data with the aim of minimizing a loss function. The learned filters are a combination of different (mainly band-pass and band-reject) filters with selective attenuation levels for different frequency levels. The filters of the first layers (CL1 and CL2) do not exhibit dominant frequencies and are quite noisy. On the other hand, the filters learned at the deeper layers (CL3 and CL4) are more regular filters, i.e., they have a well-defined frequency response which is closer to ideal filters. However, the resolution of the Fourier transform of the deeper layers is lower than in the initial layers because they are smaller than the initial ones. This analysis lead us to propose some enhancements to the proposed approach as an attempt to improve the response of the filters learned by the network.

\subsection{Architecture Enhancement}
\label{sec:arch_enh}
Three enhancements to the proposed approach are evaluated: (i) replacing the Hamming sliding window by a rectangular window because the Hamming window smooths the signal and reduces the energy of the beginning and end of the audio frame and this may cause a loss of information; (ii) augmenting slightly the amount of training data by increasing the window overlapping during the audio segmentation; (iii) initializing the first convolutional layer as a Gammatone filterbank as described in Section \ref{sec:arch}, and make this layer non-trainable.  

Table \ref{table:acc_gamm_org} summarizes the three proposed enhancements and their impact on the mean accuracy. The rectangular window leads to a slight improvement of 2\% in the mean accuracy. Increasing the overlapping from 50\% to 75\% led to another 2\% of improvement in the mean accuracy. Finally, initializing the first layer of such a 1D CNN with a Gammatone filterbank, also contributed to improve the mean accuracy in 2\%, even if the number of parameters doubles due to increase of the number of filters in such a layer. An important remark is that all these enhancements have also improved the performance of most of the other 1D CNN architectures presented in Table \ref{tab:arch}. In spite of that, the 16,000-input 1D CNN remains the one with the highest mean accuracy.

\begin{table*}[htpb!]
\caption{Improvements in the mean accuracy for the 16,000-input 1D CNN on the UrbanSound8k dataset.} 
\centering 
\footnotesize
\begin{tabular}{l l c l c l}
\hline\hline 
\textbf{CL1} & \textbf{Window} & \textbf{Overlapping} & \textbf{Combination} & \textbf{Mean} & \textbf{\# of }  \\
\textbf{Initialization} &  &  & \textbf{Rule} & \textbf{Accuracy} & \textbf{Parameters} \\ \hline
Randomly  & Hamming            & 50\%                      & Sum Rule                                & 83\% & 256,538          \\
Randomly & Rectangular              & 50\%                      & Sum Rule                                & 85\% & 256,538          \\
Randomly & Rectangular              & 75\%                      & Sum Rule                                & 87\% &  256,538         \\
Gammatone  & Rectangular             & 50\%                      & Sum Rule                                & \textbf{89\%} &  550,506        \\ \hline
\end{tabular}
\label{table:acc_gamm_org}
\end{table*}

\begin{figure*}[htpb!]
  \centering
  \includegraphics[width=1\textwidth]{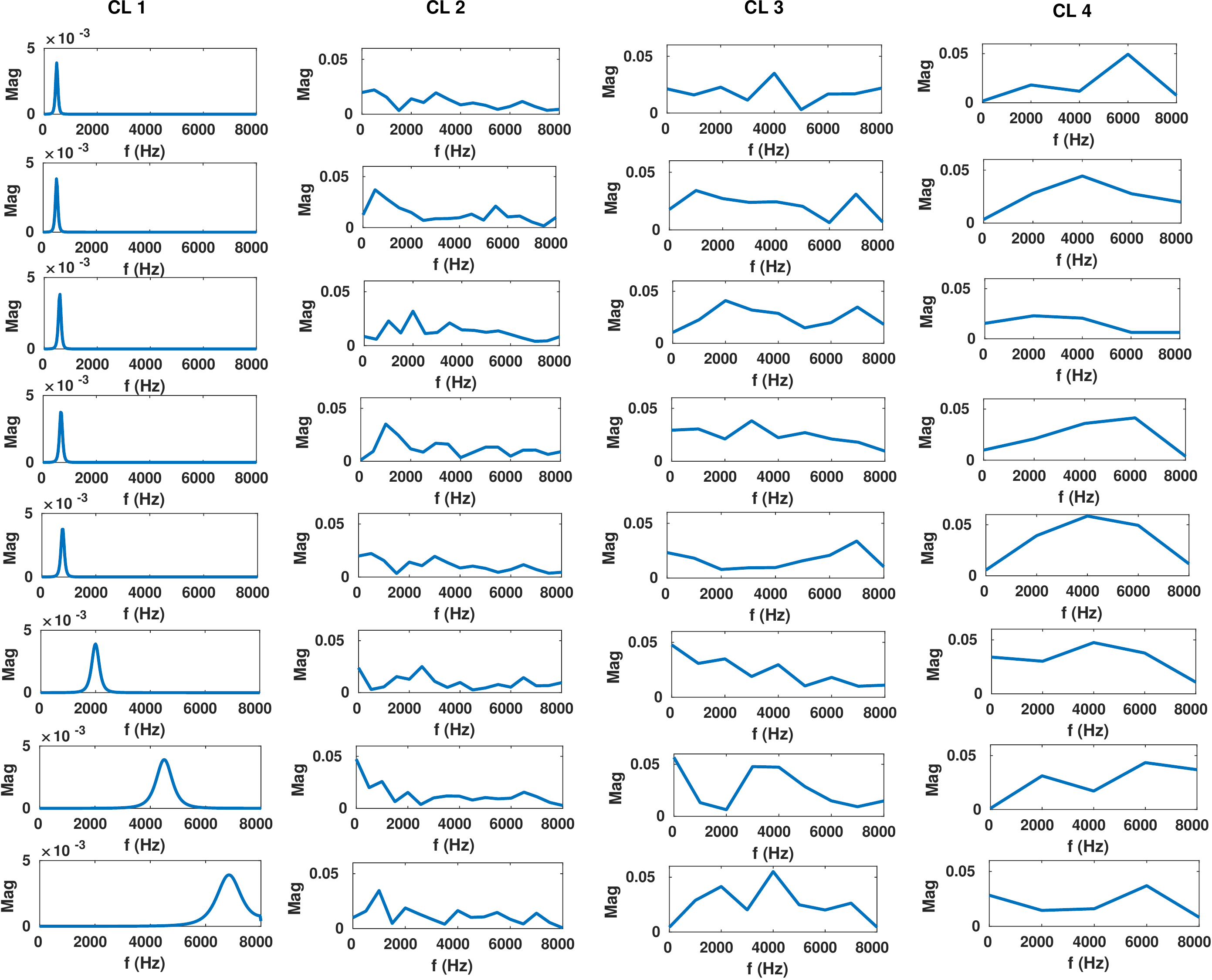}
  \caption{Fourier transform of randomly selected filters from the four convolutional layers (CLs) of the 16,000-input 1D CNN with Gammatone filterbank in the first convolutional layer of the network.}
  \label{fig:kernelsgamma}
\end{figure*}

Figure \ref{fig:kernelsgamma} shows the Fourier transform of some of the filters of the enhanced model with non-trainable Gammatone filterbank. Similar to the filters of the original model (Figure \ref{fig:kernels}), the filters of the deepest layers (CL3 and CL4) have a well-defined frequency response. Filters of the intermediate layer (CL2) still do not exhibit dominant frequency levels. Even thought, the minor changes in the responses of the intermediate and deeper filters, the Gammatone filters of the first layer were useful to improve the mean accuracy of the proposed 1D CNN.   


\section{Discussion}
\label{sec:dis}

\begin{figure}[htpb!]
  \centering
  \centerline{\includegraphics[width=9.5cm, height=9.5cm]{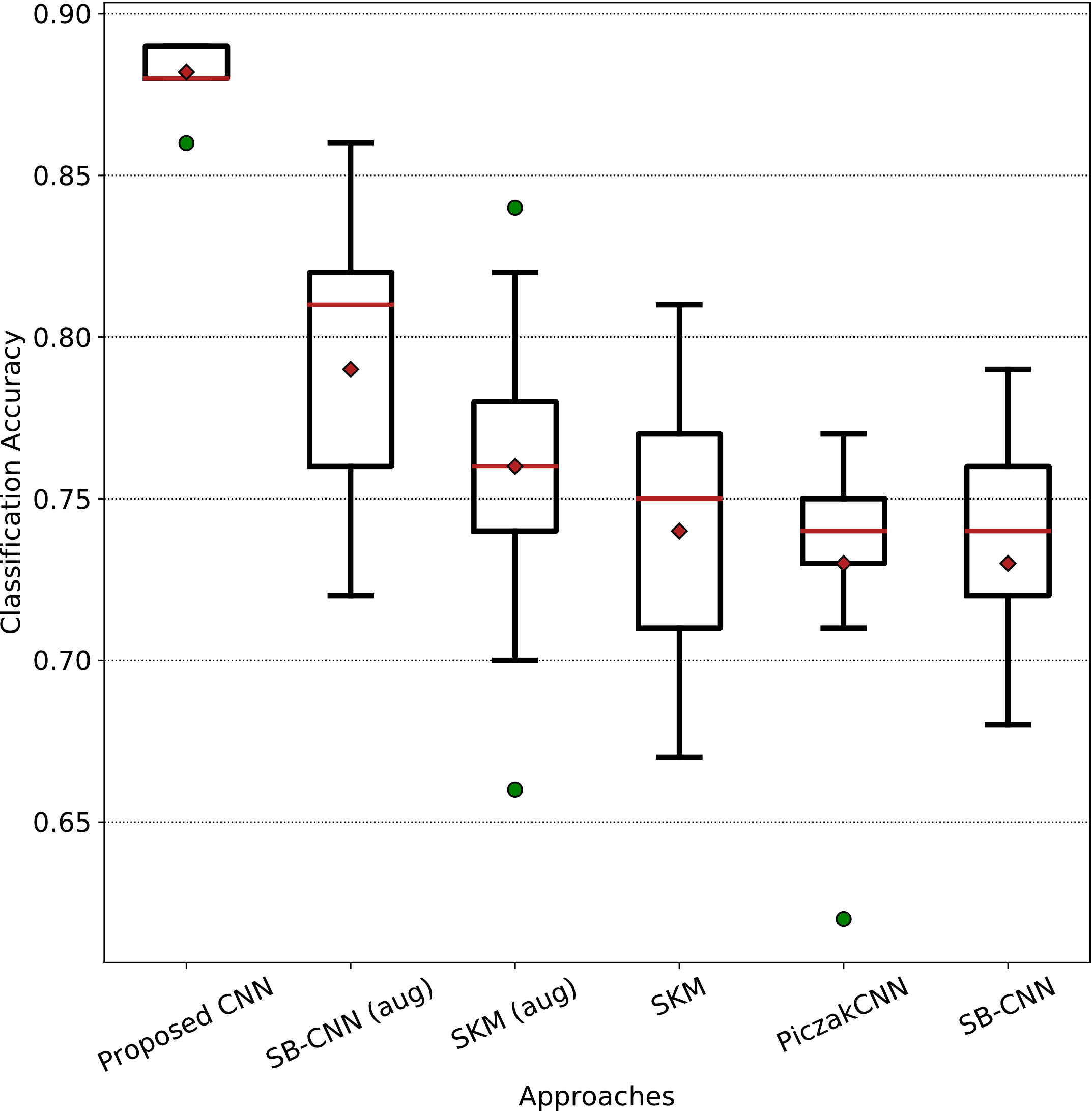}}
\caption{Classification accuracy of the proposed 1D CNN as well as the results obtained by other state-of-the-art approaches. Figure adapted from \citep{2017deepsalamon}.}
\label{fig:boxall}
\end{figure}

Table \ref{table:meanacc} shows the mean classification accuracy achieved by the proposed 1D CNN as well as the results achieved by other state-of-the-art approaches described in the literature. The proposed 1D CNN achieved a mean accuracy of 89\% with a standard deviation of only $0.9\%$ across the 10 folds. Note that the VGG architecture \citep{pons2018randomly} is implemented based on non-trained (randomly weighted) method for feature extraction. The proposed 1D CNN, the EnvNet-v2 \citep{tokozume2017learning} and the M18 CNN \citep{Dai2017} are end-to-end architectures, which learn the representation directly from the audio waveform while all other approaches in Table \ref{table:meanacc} use 2D representations of the audio signal as input. Therefore, besides the reduced number of parameters of the proposed end-to-end 1D CNN, which indicates that the proposed approach can be trained with fewer number of training samples relative to most of the other approaches, it outperforms all other approaches. Figure \ref{fig:boxall} compares the proposed 1D CNN with other approaches reported in \citep{salamon2015unsupervised} using a boxplot generated from the accuracy scores of 10 folds. We can see that the proposed 1D CNN has the highest mean and median, the interquartile range is one of the smallest one and there is no outlier. 

\begin{table}[htpb!]
\caption{Mean accuracy of different approaches on the UrbanSound8k dataset.} 
\centering 
\footnotesize
\begin{tabular}{l c c l} 
\hline\hline 
\textbf{Approach}& \textbf{Representation} & \textbf{Mean} &\textbf{\# of} \\
 &  & \textbf{Accuracy} &\textbf{Parameters} \\
\hline 
\textbf{Proposed 1D CNN Gamma}& 1D &\textbf{89\%} & 550 k \\
Proposed 1D CNN Rand & 1D & 87\% & 256 k \\
SB-CNN (DA) \citep{2017deepsalamon}& 2D & 79\% & 241 k\\ 
EnvNet-v2 \citep{tokozume2017learning}&1D &78\% & 101 M\\
SKM (DA) \citep{salamon2015unsupervised}&2D &76\% & NA\\ 
SKM \citep{salamon2015unsupervised}&2D &74\% & NA\\
PiczakCNN \citep{piczak2015environmental}&2D &73\% & 26 M\\ 
M18 CNN \citep{Dai2017}&1D &72\% & 3.7 M\\ 
SB-CNN \citep{2017deepsalamon}&1D &73\% & 241 k\\
VGG \citep{pons2018randomly}&2D &70\% & 77 M\\ 
\hline 
\multicolumn{3}{l}{NA: Not available. DA: With data augmentation.}
\end{tabular}
\label{table:meanacc} 
\end{table}

\begin{figure}[htpb!]
\begin{minipage}[b]{1.0\linewidth}
  \centering
  \centerline{\includegraphics[width=7cm, height=7cm]{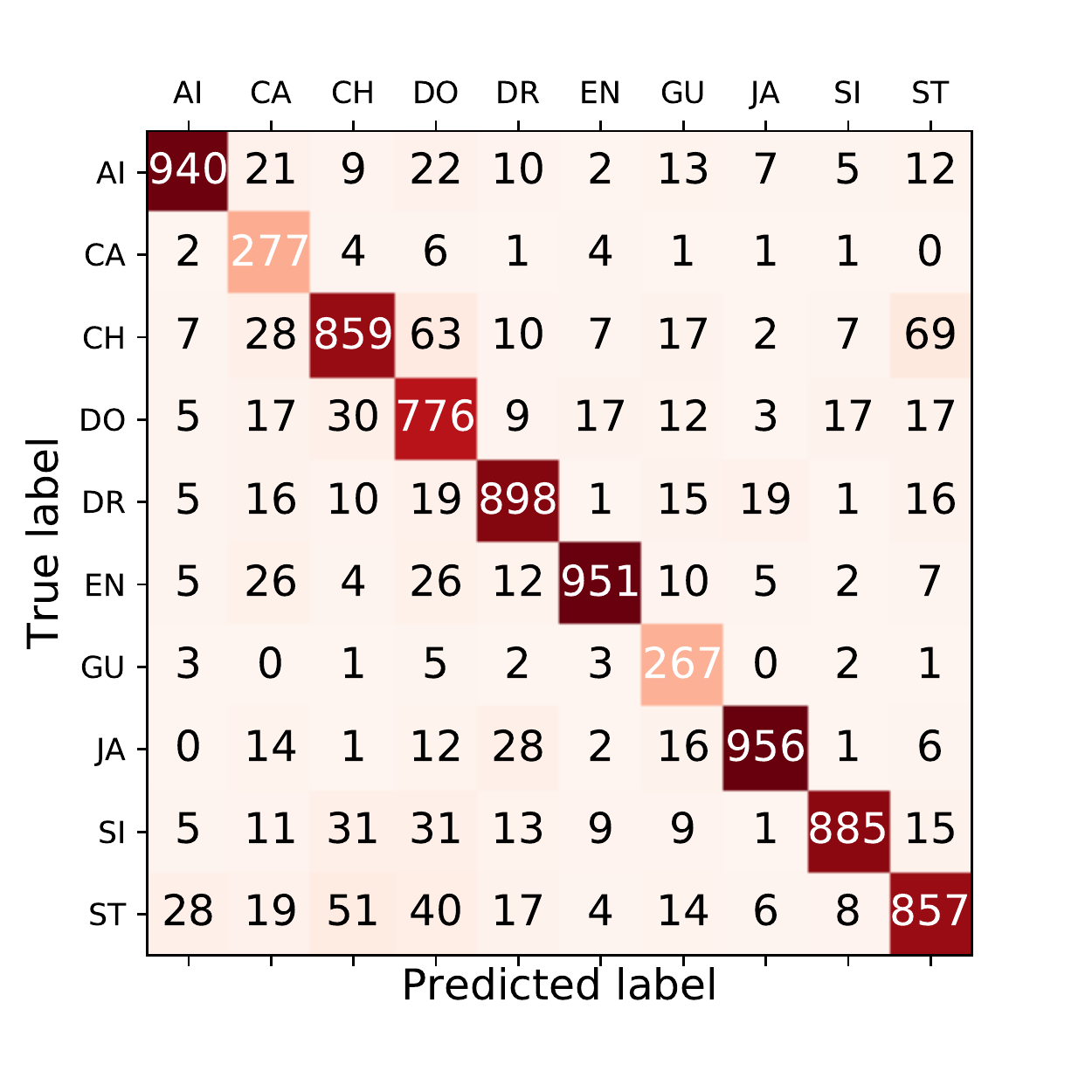}}
\end{minipage}
\caption{Confusion matrix for the proposed end-to-end 1D CNN.}
\label{fig:confusion}
\end{figure}
\begin{figure}[htpb!]
\begin{minipage}[b]{1.0\linewidth}
  \centering
  \centerline{\includegraphics[width=0.9\textwidth]{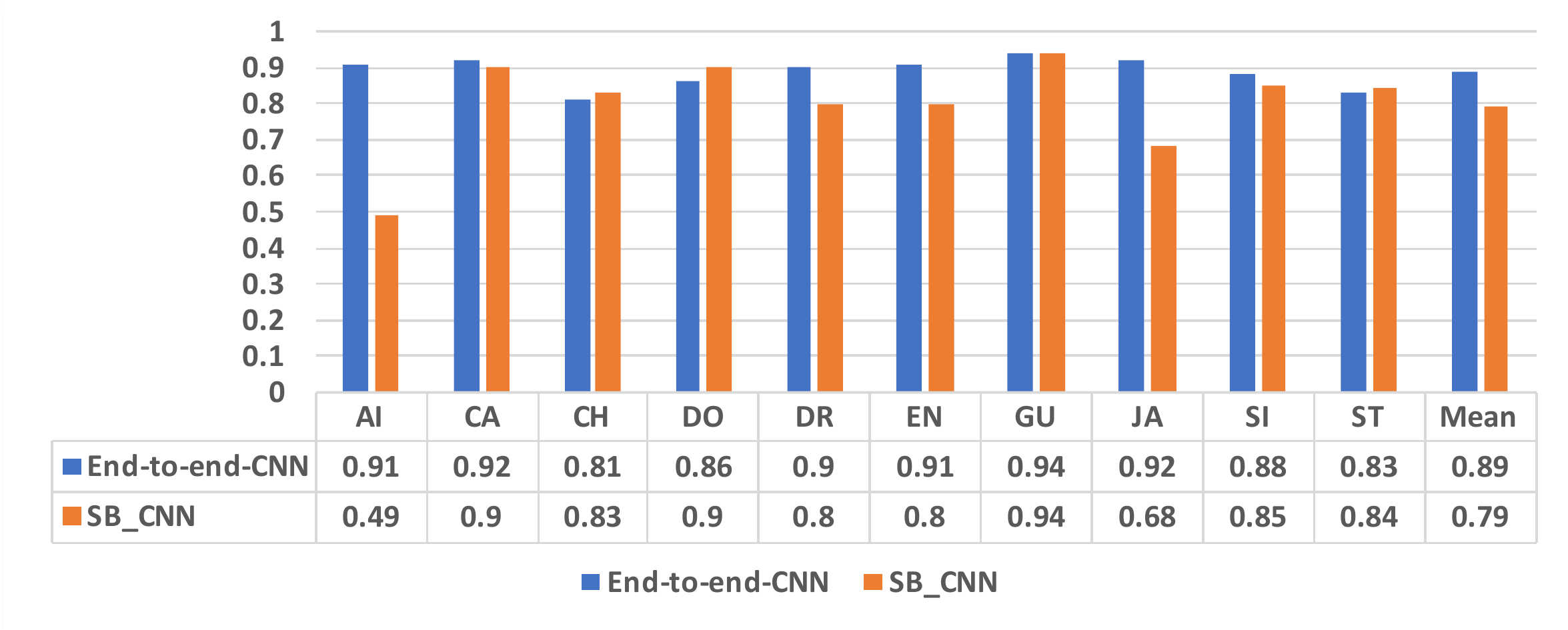}}
\end{minipage}
\caption{Per-class weighted accuracy for the proposed 1D CNN and the SB-CNN with data augmentation \citep{2017deepsalamon}.}
\label{fig:comparison}
\end{figure}

Figure \ref{fig:confusion} shows the confusion matrix of the proposed end-to-end 1D CNN on the UrbanSound8k dataset. Values along the diagonal indicate the number of samples classified correctly for each specific class. It shows that the ST and CH classes are the hardest classes for the CNN. However, EN and GU classes are well separated by the proposed CNN. For a better insight into the performance of the proposed end-to-end CNN, Figure \ref{fig:comparison} shows the per-class accuracy for each specific class in the UrbanSound8k dataset in comparison with the SB-CNN with data augmentation which uses a 2D representation \citep{2017deepsalamon}. Figure \ref{fig:comparison} shows that for class GU and ST the proposed CNN achieves relatively similar result that the SB-CNN. For classes AI, DR, EN, SI, CA and JA, the proposed CNN works better than the SB-CNN while for classes CH and DO the SB-CNN approach is slightly better than the proposed end-to-end CNN. Since the input of the proposed CNN is the audio waveform, it seems that the network tends to extract different features from the audio waveform which might be missing in the spectrogram representation of the signal. However, the SB-CNN, which uses spectrograms as input \citep{2017deepsalamon}, needs 20 times more samples than the proposed end-to-end 1D CNN to achieve the accuracy of 79\%. It is interesting to note that both classifiers could be fused in order to improve the overall accuracy since it can be seen that these approaches have different per-class accuracy for some classes such as AI and JA.

\subsection{Filter Response}
\label{sec:filred}
\begin{figure*}[htpb!]%
\centering
\subfigure[]{%
\includegraphics[width=.22\textwidth]{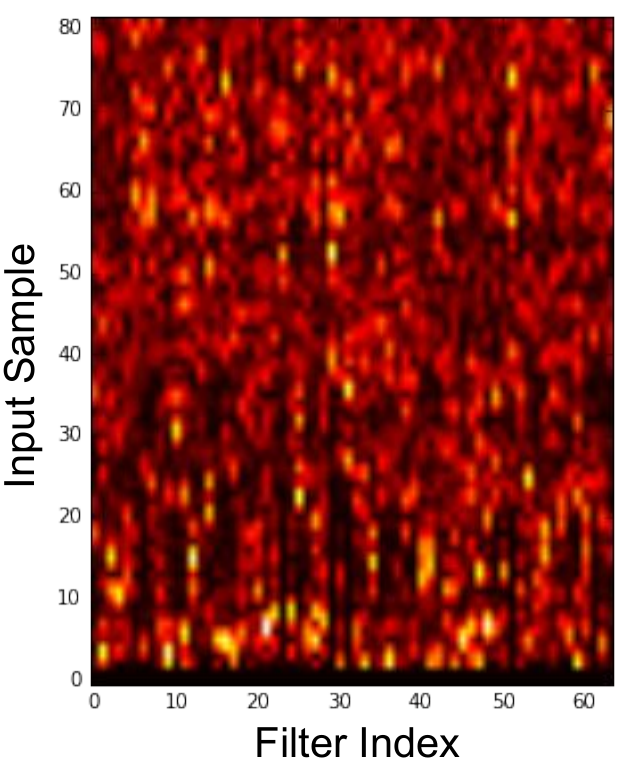}}%
\subfigure[]{%
\includegraphics[width=.22\textwidth]{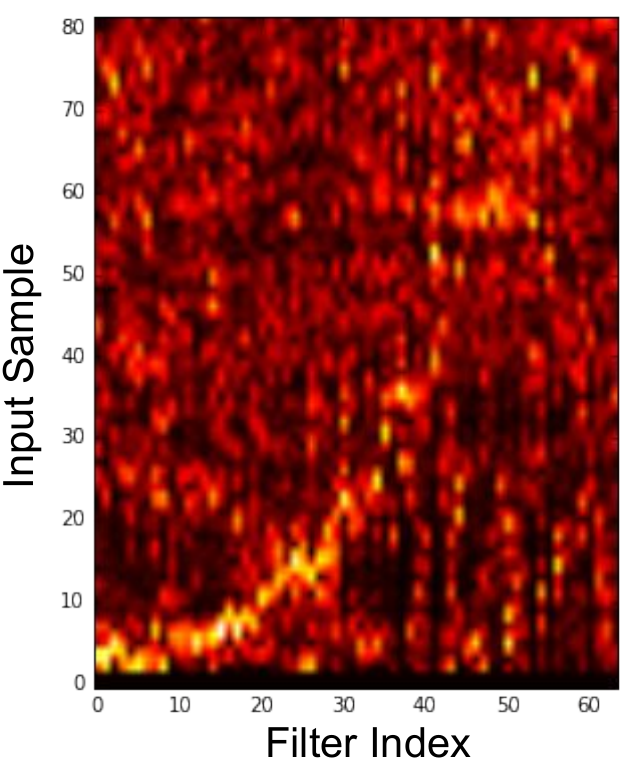}}%
\subfigure[]{%
\includegraphics[width=.22\textwidth]{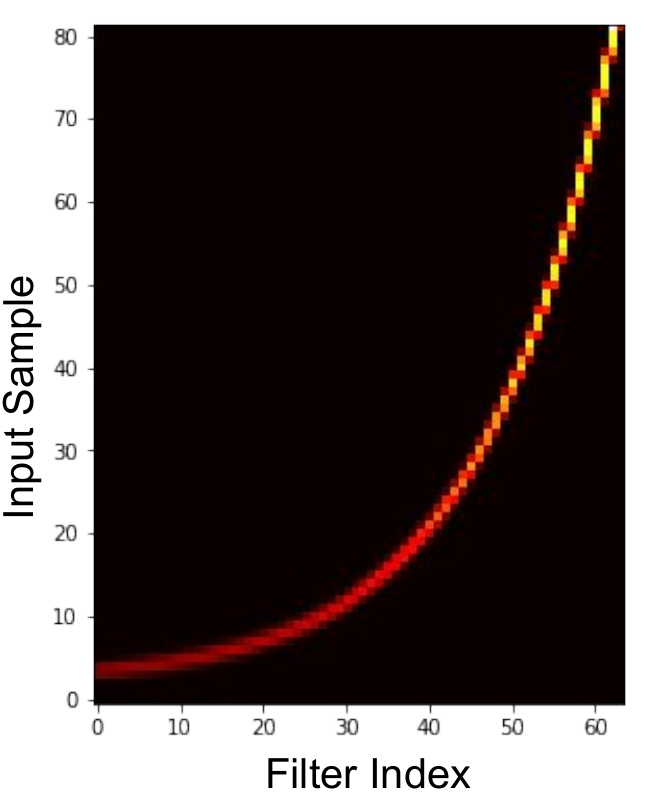}}%
\caption{Magnitude response of the convolutional filters of the first layer of the end-to-end 1D CNN: (a) response of the filters after convergence; (b) response of the kernels after sorting based on their central frequency; (c) frequency response of band pass filters. Center frequency of filters are selected according to constant $Q$ transform rules \citep{brown1991calculation}.}
\label{fig:magres}
\end{figure*}

The magnitude responses of the convolutional filters of the first layer of the proposed 1D CNN are shown in Figure \ref{fig:magres}. To obtain a better image representation of the frequency response, the number of kernels in the first layer has been increased to 64 (compared to 16 in the one used in the experiments). Note that this configuration led to a slight decrease in the classification accuracy. Figures \ref{fig:magres}(a) and \ref{fig:magres}(b) show the response of the filters after convergence and the response of the kernels sorted based on their central frequencies, respectively. The central frequency of each kernel is computed by computing the Fast Fourier transform of the filter and by selecting the frequency bin with the highest peak. Each row in the image is created by feeding the network with a sinusoidal wave with a specific frequency. For such an aim, sinusoidal waves in the range of 1 Hz to 8 kHz, with a step of 100 Hz, have been used. The feature map of the first convolutional layer is first obtained and then, it is computed the average of the feature map along the time axis. Figure \ref{fig:magres}(c) shows the output of 64 Gammatone filters used as band-pass filters.

From Figure \ref{fig:magres}(b), it can be seen that the learned filters have a logarithmic response similar to the band pass filters created using cardinal sinusoidal functions. In addition, this behavior is also similar to how humans perceive sounds, which is also logarithmic \citep{roederer2008physics}. A similar behavior has also been observed in other end-to-end systems for audio processing tasks \citep{hoshen2015speech, sainath2015learning, tokozumeENV}.

\section{Conclusion}
\label{sec:conc}
In this paper, an end-to-end 1D CNN for environmental sound classification has been proposed. The architecture of the network consists of three to five convolutional layers, depending on the length of the audio signal. Instead of using handcrafted static filterbanks such as those used to extract MFCC features, the proposed 1D CNN learns the filters directly from the audio waveform. The proposed approach was evaluated on a dataset of 8,732 audio samples and the experimental results have shown that the proposed end-to-end approach learns several relevant filter representations which allows it to outperform other state-of-the-art approaches based on 2D representations and 2D CNNs. Furthermore, the proposed end-to-end 1D architecture has fewer parameters than most of the other CNN architectures for environmental sound classification, while achieving mean accuracy that is between 11.24\% and 27.14\% higher than such 2D architectures.   

However, even if we have achieved the best results using 1D representation of the audio signal, it may have a complementarity between the learned 1D filters and the filters learned from 2D representations (spectrograms), at least for some classes, as highlighted in Figure \ref{fig:comparison}. This is an indication that the overall performance may be improved by combining the approaches that use 1D and 2D representations. As a future work, we will investigate if such a combination is feasible and if it can lead to a better performance in classifying environmental sounds. Furthermore, the filters learned in the intermediate convolutional layers of the proposed 1D CNN do not exhibit dominant frequencies and seems to be noisy. A further investigation is necessary to find out how to circumvent such a problem and possibly improve further the performance of the proposed 1D CNN.

\section*{Availability of Data and Material}
UrbanSound8k dataset \citep{Salamon:2014:DTU:2647868.2655045} is used for training and testing the method. The dataset is available online at: https://urbansounddataset.weebly.com/urbansound8k.html.

The source code of the proposed end-to-end CNN will also be made available in the final version of the paper.

\section*{Competing Interests}
The authors declare that they have no competing interests.

\section*{Credit Authorship Contribution Statement}
\textbf{Sajjad Abdoli}: Conceptualization, Methodology, Software, Validation, Formal Analysis, Investigation, Data Curation, Writing – Original Draft, Visualization. \textbf{Patrick Cardinal and Alessandro Lameiras Koerich}: Conceptualization, Methodology, Validation, Formal Analysis, Investigation, Resources, Data Curation, Writing – Review \& Editing, Supervision, Project Administration, Funding Acquisition. 

\section*{Acknowledgements}
 This work was funded by the Natural Sciences and Engineering Research Council of Canada (NSERC). This work was also supported by the NVIDIA GPU Grant Program.

\section*{References}

\bibliography{mybibfile}

\end{document}